\documentclass[11pt]{article}
\usepackage{amssymb,amsopn,amsthm}

\advance\textwidth by 10mm
\advance\oddsidemargin by -5mm
\advance\evensidemargin by -5mm

\DeclareMathOperator{\tr}{tr}
\DeclareMathOperator{\supp}{supp}
\DeclareMathOperator{\image}{im}
\DeclareMathOperator{\Irr}{Irr}
\newcommand{\onemat}[0]{{\mathbf 1}}

\newtheorem{theorem}{Theorem}
\newtheorem{lemma}[theorem]{Lemma}
\newtheorem{corollary}[theorem]{Corollary}

\newtheorem{example}[theorem]{Example}

\newcommand{\nix}[1]{}
\newcommand{\C}{\mathbf{C}}

\makeatletter
\renewcommand{\@maketitle}{%
  \newpage
  \null
  \vskip 2em%
  \begin{center}%
  \let \footnote \thanks
    {\large \@title \par}%
    \vskip 1.5em%
    {\normalsize
      \lineskip .5em%
      \begin{tabular}[t]{c}%
        \@author
      \end{tabular}\par}%
    \vskip 1em%
    {\large \@date}%
  \end{center}%
  \par
  \vskip 1.5em}
\renewcommand\section{\@startsection {section}{1}{\z@}%
                                   {-3.5ex \@plus -1ex \@minus -.2ex}%
                                   {2.3ex \@plus.2ex}%
                                   {\normalfont\large\bfseries}}
\makeatother

\newenvironment{keywords}{\small\textbf{Keywords}}{\medskip}

\begin{document}

\title{\Large\textbf{Remarks on Clifford Codes}}
\author{
Andreas Klappenecker${}^{*}$, Martin R\"otteler${}^{\dagger}$\\[1ex]
${}^*$Department of Computer Science, Texas A\&M University,\\
College Station, TX 77843-3112, USA,\\
\texttt{klappi@cs.tamu.edu}\\
${}^\dagger$Institute for Quantum Computation, University of Waterloo\\ 
Waterloo, Ontario, Canada N2L 3G1,\\
\texttt{mroetteler@iqc.ca}
}

\date{}
\maketitle

\begin{abstract}
\noindent Clifford codes are a class of quantum error control codes
that form a natural generalization of stabilizer codes.  These codes
were introduced in 1996 by Knill, but only a single Clifford code was
known, which is not already a stabilizer code. We derive a necessary and
sufficient condition that allows to decide when a Clifford code is a
stabilizer code, and compile a table of all true Clifford
codes for error groups of small order. 
\end{abstract}


\section{Introduction} 
Storing and manipulating the states of quantum mechanical systems for
the purpose of quantum information processing is appealing. One can
take advantage of quantum mechanical states to gain
considerable computational advantages. However, this kind of
information processing is highly sensitive to noise, and 
protection by some means of quantum error correction is necessary. 

The most common construction of quantum error correcting codes is
based on binary stabilizer codes, a concept developed by
Gottesman~\cite{gottesman96} and Calderbank, Rains, Shor, and
Sloane~\cite{calderbank98}, and others. Subsequently, stabilizer codes
were generalized to nonbinary alphabets by Ashikhmin and
Knill~\cite{ashikhmin01}, Bierbrauer and Edel~\cite{bierbrauer00},
Feng~\cite{feng02}, and Matsumoto and Uyematsu~\cite{matsumoto00}. 

A further generalization of stabilizer codes, the so-called Clifford
codes, was introduced by Knill~\cite{knill96b}. Knill realized that
it is possible to lift the restriction to \emph{abelian} normal
subgroups in the definition of stabilizer codes, allowing arbitrary
normal subgroups instead.

A fundamental problem is to decide when a Clifford code is not equal
to a stabilizer code. So far, this has been demonstrated for just a
single example~\cite{klappenecker031}. The construction of this
example required an exhaustive search of all codes based on a given
error group, because more convenient criteria did not exist at the
time.  The search for such examples is complicated by the fact that in
many error groups there simply do not exist any Clifford codes that
are not equal to stabilizer codes.

We further develop the theory of Clifford codes. We
derive a necessary and sufficient condition that allows to decide when
a Clifford code is equal to a stabilizer code. This condition enables
us to compile a table of all Clifford codes that are not
stabilizer codes for error groups of order up to 255.

\section{Basic Notions} 
Let $G$ be a finite group having an irreducible, faithful character
$\phi$ of degree $d:=\phi(1)=(G\colon Z(G))^{1/2}$. Denote by $\rho$ a
unitary representation of $G$ affording~$\phi$. The matrices
${\mathcal{E}}=\{\rho(g)\,|\, g\in G\}$ represent a discrete set of
errors. For instance, in the special case of binary stabilizer codes,
the error group $G$ is given by an extra-special $2$-group and the
representing matrices $\rho(g)$ by tensor products of Pauli matrices.

Let $N\trianglelefteq G$ be a normal subgroup of $G$, and let $\chi$
an irreducible character of $N$ such that $[\chi,\phi_N]\neq 0$.  A
\textbf{Clifford code} with data $(G,\rho,N,\chi)$ is defined to be
the image of the orthogonal projector
\begin{equation}\label{eq:clifford}
P = \frac{\chi(1)}{|N|}\sum_{n\in N} \chi(n^{-1})\rho(n),
\end{equation}
that is, a subspace of dimension $\tr P$ of $\C^d$. 
If the normal subgroup $N$
is abelian, then the Clifford code is called a \textbf{stabilizer code}.  

A fundamental problem is to decide when a Clifford code is a
stabilizer code. If $N$ is not abelian, then the image of the
projector $P$ might still be a stabilizer code, but with respect to
another normal abelian subgroup $A$ of $G$. We
derive necessary and sufficient criteria when a Clifford code is a
stabilizer code. Note that we keep the discrete set of errors fixed
for this comparison.

\section{Clifford Codes versus Stabilizer Codes} 
Let $Q\subseteq \C^d$ be a Clifford code with data $(G,\rho,N,\chi)$. 
The inertia subgroup 
$$T=\{ g\in G\,|\, \chi(n)=\chi(gng^{-1}) \mbox{ for all } n\in N\}$$
consists of all elements $g$ of $G$ such that $\rho(g)Q=Q$.  Let
$\vartheta\in \Irr(T)$ denote the character such that
$[\vartheta_N,\chi]\neq 0$ and $[\phi_T,\vartheta]\neq 0$. This is the
character afforded by the irreducible $\C T$-module $Q$. 
The quasikernel 
$$ Z(\vartheta) = \{ g\in G\,|\, \vartheta(1)=|\vartheta(g)|\}$$
consists of the elements $g$ of $G$ that act on the code $Q$ by scalar
multiplication.  These two groups characterize the errors in $G$ that
are detectable by the code. An error $\rho(g)$ is detectable by the
code $Q$ if and only if $g\not\in T-Z(\vartheta)$,
see~\cite{knill96b,klappenecker033}.

The group $Z(\vartheta)$ will allow us to decide whether or not the
Clifford code $Q$ is a stabilizer code. Denote by $\mathcal{A} = \{ A
\le Z(\vartheta)\,|\, A \trianglelefteq G, A \mbox{ abelian}\}$ the
set of all normal abelian subgroups $A$ of $G$ that are contained in
$Z(\vartheta)$. We will show that in case $Q$ is a stabilizer code,
then its stabilizer can be defined in terms of a maximal group of
$\mathcal{A}$.

\begin{lemma}\label{propA}
If $A\in \mathcal{A}$, then there exists a linear character $\theta$ of $A$ 
such that the image of the orthogonal projector 
\begin{equation}\label{eq:stab}
P_A = \frac{1}{|A|} \sum_{a\in A} \theta(a^{-1})\rho(a)
\end{equation}
contains $Q$, meaning that $P_av=v$ holds for all $v\in Q$. 
\end{lemma}
\begin{proof}
The condition $A\le Z(\vartheta)$ ensures that each matrix $\rho(a)$,
$a\in A$, acts by multiplication with a scalar $\theta(a)$ on the code
space $Q$. The eigenvalues $\theta(a)$ form a linear character of $A$. 
Choosing this character in~(\ref{eq:stab}) yields 
$P_Av=v$ because $\theta(a^{-1})\theta(a)=1$.
\end{proof}

\begin{lemma}
Let $A$ be an abelian normal subgroup of $G$ with linear
character $\theta$. If the image of the projector (\ref{eq:stab}) 
contains the Clifford code $Q$, then $A\le Z(\vartheta)$. 
\end{lemma}
\begin{proof}
Direct calculation shows that $\rho(a)P_A=\theta(a)P_A$ for all $a\in A$.  
Thus, $\rho(a)$ acts by
multiplication with a scalar $\theta(a)$ on the code space $Q$. It follows that $A\le Z(\vartheta)$.
\end{proof}

\begin{theorem}\label{th:cond}
Let $Q$ be a Clifford code with data $(G,\rho,N,\chi)$, and denote by
$\phi$ the irreducible character of $G$ afforded by the representation
$\rho$. Keeping the above notations, we can conclude that $Q$ is a
stabilizer code if and only if\/ $\dim Q=|A \cap Z(G)|\phi(1)/|A|$
holds for some $A\in \mathcal{A}$.
\end{theorem}
\begin{proof}
By the previous two lemmas, the image 
of the projector (\ref{eq:stab}) 
contains $Q$ if and only if $A\in \mathcal{A}$.  
It follows that $\image P_A=Q$ if and only if 
$\dim(\image P_A)=\tr P_A = |A\cap Z(G)|\phi(1)/|A|$ 
coincides with $\dim Q$. 
\end{proof}
We can strengthen this result by observing that we can assume without
loss of generality that the elements of the center $Z(G)=\{ z\,|\, zg=gz
\mbox{ for all } G\}$ of $G$ are contained in the normal subgroup $N$.

\begin{lemma}
Let $Q$ be a Clifford code with data $(G,\rho,N,\chi)$.  
It is possible to define $Q$ over the normal subgroup $N_Z=NZ(G)$ of $G$ that is obtained by extending $N$ by central elements of $G$.
\end{lemma}
\begin{proof}
Let $\sigma$ denote a unitary representation of $N$ with character
$\chi$.  We can extend this representation to a representation
$\sigma_Z$ of $N_Z$ by defining $\sigma_Z(zn)=\alpha(z)\sigma(n)$,
where $\alpha(z)$ is the scalar $\rho(z)=\alpha(z)\mathbf{1}$, $z\in
Z(G)$.  It is easy to check that this is well-defined, because
$\sigma$ is a constituent of the restriction of the representation
$\rho$ to $N$. Denote by $\chi_Z$ the character $\chi_Z=\tr \sigma_Z$.
Let $Z$ be a transversal of $N_Z/N$ such that $Z\subseteq Z(G)$. 
Then the definition of the character $\chi_Z$ ensures that 
$$ \frac{\chi_Z(1)}{|N_Z|} \sum_{z\in Z}\sum_{n\in N_Z} \chi(n^{-1}z^{-1})\rho(zn) = \frac{\chi_Z(1)}{|N_Z|} \sum_{z\in Z}\sum_{n\in N_Z} \chi(n^{-1})\rho(n)
$$  
coincides with the projector onto the Clifford code Q given by (\ref{eq:clifford}); therefore 
$(G,\rho,N_Z,\chi_Z)$ defines the Clifford code $Q$. 
\end{proof}

From now on, we will assume without loss of generality that the center
$Z(G)$ of $G$ is contained in $N$. If $Q$ is not a stabilizer code, then we
can prove this by looking at just one group in $\mathcal{A}$ that is
of maximal order:

\begin{corollary}
A Clifford code~$Q$ is a stabilizer code if and only if\/ $\dim
Q=|Z(G)|\phi(1)/|A|$ holds for a group $A\in \mathcal{A}$ that has
maximal order among the groups in $\mathcal{A}$.
\end{corollary}
\begin{proof}
If $Q$ is a stabilizer code, then there exists some $A\in\mathcal{A}$
such that $Q$ is the image of the projector~(\ref{eq:stab}).  By the
previous lemma, we can assume without loss of generality that $Z(G)$
is contained in $A$, hence the dimension of the code is $\tr P_A
=|Z(G)|\phi(1)/|A|=\dim Q$. Seeking a contradiction, we assume that
this group $A$ is not a group of maximal order in $\mathcal{A}$.  This
would imply that $\mathcal{A}$ contains a group $A_*\ge Z(G)$ with
$|A_*|>|A|$. By definition of $\mathcal{A}$, the image of the projector
$P_{A_*}$ contains $Q$, hence
$$ \dim Q< |Z(G)|\phi(1)/|A_*|< |Z(G)|\phi(1)/|A|=\dim Q,$$ 
yields the desired contradiction

Conversely, if $A$ is a group of maximal order in $\mathcal{A}$, then $Z(G)$ is
contained in~$A$. Therefore, $|Z(G)|\phi(1)/|A|=\dim Q$ is the
dimension of its stabilizer, and since the image of $P_A$ contains
$Q$, it follows that~$\image P_A=Q$; hence, $Q$ is a stabilizer code.
\end{proof}

If we want to show that $Q$ is a stabilizer code, then the following
condition is convenient.

\begin{corollary}\label{criterion}
Suppose that $Z(G)\le N$. 
A Clifford code
$(G,\rho,N,\chi)$ is a stabilizer code if and
only if $\chi^2(1)=|N|/|A|$ for some $A\in \mathcal{A}$ with $Z(G)\le A$. 
\end{corollary}
\begin{proof}
Compare $\tr P=\chi(1)^2\phi(1)|Z(G)|/|N|$ and $\tr
P_A=\phi(1)|Z(G)|/|A|$ and apply Theorem~\ref{th:cond}.
\end{proof}

\section{Examples} 
We give two examples to illustrate the method. In the first example, we
construct a Clifford code for the data $(G, \rho, N, \chi)$ which
turns out to be equal to a stabilizer code for an abelian normal
subgroup $A \lhd G$. In the second example, we construct 
a true Clifford code, that is, we rule out that this code is equal to a
stabilizer code by Corollary~\ref{criterion}. This shows that the
class of Clifford codes is strictly bigger than the class of
stabilizer codes.

\begin{example}\rm 
Let $G$ be a group. Recall that the commutator of the elements 
$g,h \in G$ is defined by $[g,h] :=g^{-1} h^{-1} g h$. Now, 
consider the group $G$ given by the power-commutator presentation
\begin{eqnarray*}
G & = & \langle a, b, c, d, e | \; a^2 = d, \; b^2, \; c^2, \; d^2, \;
  e^2, \; [b,a]=c, \; [c,a]=e, \; [d,a], \; [e,a] \\
& & \phantom{\langle a, b, c, d, e | \; } [c, b], \; [d,b]= e, \;
  [e,b], \; [d, c], \; [e, c], \; [e,d] \rangle.
\end{eqnarray*}
We find that $G$ is a finite group of size $32$. The group $G$ has an
irreducible four-dimensional representation $\rho$. Since the order of
the centre is $|Z(G)|=2$ we have that $[G:Z(G)]={\rm deg}(\rho)^2$,
i.\,e., $G$ is an error group. Actually, $G$ is isomorphic to the
group \texttt{SmallGroup(32,6)} which is contained in the catalog of
error groups of small order \cite{klappenecker034}. Also note
that the set $\{a, b, c, d, e\}$ is not a minimal generating set of
$G$. In fact, we have that $G = \langle a, ab \rangle$.  
We find that $\rho$ is the irreducible four-dimensional 
representation given by 
\[
\rho(a) = \left(
\begin{array}{rrrr}
   0 &   0 &   \phantom{-}1 &   0 \\
   0 &   0 &   0 &   \phantom{-}1 \\
   \phantom{-}1 &   0 &   0 &   0 \\
   0 &  -1 &   0 &   0 
\end{array}
\right), \quad
\rho(ab) = \left(
\begin{array}{rrrr}
   0 &   0 &   0 &  -1 \\
   0 &   0 &  -1 &   0 \\ 
   0 &   \phantom{-}1 &   0 &   0 \\ 
  -1 &   0 &   0 &   0 
\end{array}
\right).
\]

In search for suitable Clifford codes we choose the normal subgroup $N
= \langle b, c, d, e \rangle$ of order $16$. 
Then $N$ has ten conjugacy classes with representatives 
\[
 (), \; e, \; d, \; c, \; ce, \; cd, \; b, \; bd, \; bc, \; bcd.
\]
We define the irreducible character $\chi$ with respect to this
ordering of the conjugacy classes. Then $\chi$ is given by the values
\[
 2, \; -2, \; 0, \; 2, \; -2, \; 0, \; 0, \; 0, \; 0, \; 0.
\] 
Next, we construct the corresponding projector $P =
\frac{\chi(1)}{|N|}\sum_{n\in N} \chi(n^{-1})\rho(n)$, which turns out
to be 
\[ 
P =  
\left(
\begin{array}{rrrr}
   1 &   0 &   0 &   0 \\
   0 &   1 &   0 &   0 \\ 
   0 &   0 &   0 &   0 \\ 
   0 &   0 &   0 &   0 
\end{array}
\right).
\]
The corresponding Clifford code $Q$ is equal to a stabilizer
code. In order to see this, we have to compute the inertia group
$T(\chi)$ first. In turns out to be no larger than the normal
subgroup, i.\,e., we have $T(\chi) = N$. Hence the representation
$\vartheta$ of the inertia group coincides with $\chi$ in this
case. The quasikernel $Z(\chi)$ is a group of order four given by
$Z(\vartheta) = \langle c, e \rangle$. Note that $Z(\vartheta) \equiv
Z_2 \times Z_2$, i.\,e., the quasikernel is not cyclic.

We obtain that ${\rm deg}(\chi)^2 = 4 = \frac{|N|}{|Z(\vartheta)|}$,
i.\,e., by the criterion given in Corollary \ref{criterion} the code
is a stabilizer code. 

Indeed, we can construct this stabilizer code as follows: the
characters $\epsilon_1, \ldots, \epsilon_4$ of $Z(\vartheta)$ are
given by the rows of the Hadamard matrix
\[
H_2 \otimes H_2 = 
\left(
\begin{array}{rrrr}
   1 &   1 &   1 &   1 \\
   1 &  -1 &   1 &  -1 \\ 
   1 &   1 &  -1 &  -1 \\ 
   1 &  -1 &  -1 &   1 
\end{array}
\right).
\]
We find that the Clifford code corresponding to $(G, \rho,
Z(\vartheta), \epsilon_3)$ is equal to $Q$. 
\end{example}

The preceding example shows that there are Clifford codes $(G, \rho,
N, \chi)$ for which the corresponding projector is equal to a
stabilizer code $(G, \rho, A, \chi^\prime)$, where $A \lhd G$ is an
{\em abelian} subgroup of $G$. In \cite{klappenecker031} is has been
shown that there is an error group of size $32$ (namely, the group
\texttt{SmallGroup(32,5)}) which gives rise to a true Clifford
code. The dimension of the corresponding ambient space in this example
was four and the dimension of the code, i.\,e., the rank of the
projector, was two. In the following example we exhibit a true
Clifford code for a six-dimensional system.

\begin{example}\rm
Let $G$ be the group generated by the matrices $A, B, C \in
\C^{6\times 6}$ which are defined as follows:
\[
A := 
\left(
\begin{array}{rrrrrr}
  \cdot &   1 &   \cdot &   \cdot &   \cdot &   \cdot \\
  \cdot &   \cdot &   1 &   \cdot &   \cdot &   \cdot \\
  1 &   \cdot &   \cdot &   \cdot &   \cdot &   \cdot \\
  \cdot &   \cdot &   \cdot &   \cdot &   \cdot &   1 \\
  \cdot &   \cdot &   \cdot &   1 &   \cdot &   \cdot \\
  \cdot &   \cdot &   \cdot &   \cdot &   1 &   \cdot 
\end{array}
\right), 
\quad
B := 
\left(
\begin{array}{rrrrrr}
    \omega^3 &           \cdot &           \cdot &           \cdot &           \cdot &           \cdot \\
      \cdot &     \omega^7 &           \cdot &           \cdot &           \cdot &           \cdot \\
      \cdot &           \cdot &    \omega^{11} &           \cdot &           \cdot &           \cdot \\
      \cdot &           \cdot &           \cdot &       -\omega^3 &           \cdot &           \cdot \\
      \cdot &           \cdot &           \cdot &           \cdot &   -\omega^{11} &           \cdot \\
      \cdot &           \cdot &           \cdot &           \cdot &           \cdot &    -\omega^7 
\end{array}
\right), 
\]
and $C := \left( \begin{array}{rr} 0 & 1 \\ -1 & 0 \end{array} \right)
\otimes \onemat_3$. Here $\omega := \exp(2 \pi i/12)$ is a primitive
$12$th root of unity and we have replaced entries equal to $0$ with a
dot. Then $G$ is isomorphic to the group \texttt{SmallGroup(216,66)}
an is an abstract error group, i.\,e., $[G : Z(G)] = {\rm
deg}(\phi)^2$, where $\phi$ is the character of degree $6$
corresponding to the natural matrix representation of $G$.

As normal subgroup $N$ we choose the group generated by $N = \langle
A, C, D\rangle$, where $D = {\rm diag}(1, \omega_3, \omega_3^2, 1,
\omega_3^2, \omega_3)$. Here $\omega_3 := \exp(2 \pi i/3)$. We find
that $N$ has order $108$ and that $N$ has a character $\chi$ 
leading to the projector
\[
P = 
\frac{1}{2}\,\left(
\begin{array}{rrrrrr}
      1 &      \cdot &      \cdot &      \phantom{-}i &      \cdot &      \cdot \\
      \cdot &      1 &      \cdot &      \cdot &      \cdot &      \phantom{-}i \\
      \cdot &      \cdot &      1 &      \cdot &      \phantom{-}i &      \cdot \\
     -i &      \cdot &      \cdot &      1 &      \cdot &      \cdot \\
      \cdot &      \cdot &     -i &      \cdot &      1 &      \cdot \\
      \cdot &     -i &      \cdot &      \cdot &      \cdot &      1 
\end{array}
\right)
\]
of rank $3$. Like in the previous example we find that the inertia
group $T(\chi)$ coincides with $N$.

We find that the quasikernel $Z(\vartheta)$ has order $6$. Hence
we obtain that the possible quotients $\frac{|N|}{A}$, where $A$ is a
subgroup of $Z(\vartheta)$ are given by $216, 108, 72$, and
$36$, i.\,e., none of them equals ${\rm deg}(\chi)^2 = 9$. This shows
that the Clifford code $Q$ defined by the projector $P$ is not equal
to a stabilizer code for the group $G$.
\end{example}

\section{Nonstabilizer Clifford Codes of Small Order}
We have compiled a table containing information about Clifford codes
for error groups of small order to illustrate the above theory.  We
have carried out a search over all error groups up to size $255$,
where we made use of the fact that information about these groups is
available in~\cite{klappenecker034}. We maintain a catalog of error
groups of small order at 
\begin{center}
\texttt{http://faculty.cs.tamu.edu/klappi/ueb/ueb.html } 
\end{center}
from which the relevant information about error groups can be extracted.

For each error group $G$, we constructed Clifford codes for all
possible data $(G, \rho, N, \chi)$. This includes a computation of the
lattice of normal subgroups of $G$ and a search for all suitable
characters $\chi$. Using the criterion given in Corollary
\ref{criterion}, we have filtered all those Clifford codes which are
{\rm not} equivalent to a stabilizer code. In Table~\ref{cliffTable},
we give condensed information about the result of this search. 
The catalog number refers to the number in the Neub\"user
catalog used in MAGMA and GAP, cf.~\cite{magma,gap}, which can be
accessed via the command \texttt{SmallGroup(<size>, <no>)}. The
character $\phi$ is the character of $G$ corresponding to $\rho$ which
defines the error basis. The normal subgroup $N$ together with its
character $\chi$ defines the Clifford code. In the last column we give
the dimension of the resulting Clifford code. For instance the row
\begin{center}
\begin{tabular}{|c||c|c|c|c|c|}
\hline
216  & \qquad\qquad66, 68, 72\qquad\qquad  &  \quad 6 \quad  &  
\quad 108 \quad &  \quad 3 \quad  & \quad  3 \quad \\
\hline
\end{tabular}
\end{center}
shows that there are three different abstract error groups of size
$216$ which give rise to nonstabilizer Clifford codes. These groups can be 
constructed using the commands 
\begin{center}
\texttt{SmallGroup(216,66)}, \texttt{SmallGroup(216,68)}, and \texttt{SmallGroup(216,72)}
\end{center}
in GAP.  The third column shows that all these error groups give rise to
six-dimen\-sional error groups. Furthermore, in each of these groups
exists a normal subgroup $N$ of order $108$ having an irreducible
character $\chi$ of degree $3$. This in turn gives rise to a projector
onto a Clifford code of dimension $3$, which is not equal to a
stabilizer code. 

\begin{table}[htbp]
\begin{tabular}{|c||c@{\,}|c@{\,}|c@{\,}|c@{\,}|c@{\,}|}
\hline%
$\#G$ & Catalog Number & $\phi(1)$ & $\#N$ & $\chi(1)$ & ${\rm dim}(Q)$  \\
\hline\hline%
32 & 43, 44 & 4 & 16 & 2 & 2 \\
\hline
64 & 102, 111, 125, 256 & 4 & 32 & 2 & 2 \\
\hline
96 & 183, 184 & 4 & 48 & 2 & 2 \\
\hline
128 & \begin{minipage}[t]{7.8cm}
71, 72, 73, 74, 87, 88, 138, 139, 634, 635, 636, 637, 
641, 642, 643, 644, 645, 646, 647, 853, 854, 855, 859, 860, 861,
865, 866, 867, 912, 913, 922, 923, 931, 932, 933, 934, 935, 970, 
971, 1751, 1752, 1753, 1754, 1758, 1759, 1760, 1800, 1801, 2020, 2021,
2317, 2318
\end{minipage} & 8 & 32 & 2& 2 \\
\hline
128 & \begin{minipage}[t]{7.8cm}
138, 139, 144, 145, 853, 854, 855, 859, 860, 861, 865, 
866, 867, 912, 913, 922, 923, 931, 932, 933, 934, 935, 970, 971,
1751, 1752, 1753, 1754, 1758, 1759, 1760, 1800, 1801, 2020, 2021,
2317, 2318
\end{minipage} & 8 & 64 & 4 & 4 \\  
\hline
128 & 912, 913, 922, 2020, 2021, 2317, 2318 & 8 & 16 & 2 & 4 \\ 
\hline
128 & 849, 885, 903, 1687 & 4 & 64 & 2 & 2 \\ 
\hline
160& 197, 198 & 4 & 80 & 2 & 2 \\
\hline
192& 854, 863, 877, 1464 & 4 & 96 & 2 & 2 \\
\hline
216& 39, 41, 42 & 6 & 72 & 2 & 2 \\
\hline
216& 66, 68, 72 & 6 & 108 & 3 & 3 \\
\hline
224& 171, 172 & 4 & 112 & 2 & 2 \\
\hline
243& 28, 29, 30, 56, 57, 58, 59, 60 & 9 & 81 & 3 & 3 \\
\hline
\end{tabular}
\caption{\label{cliffTable} Error groups of size up to $255$ giving
rise to Clifford codes which are not equivalent to stabilizer codes. Note
that some groups be listed in several rows of this table since. This
is due to the fact that Clifford codes for various normal subgroups
$N$ and characters $\chi$ might exist.}
\end{table}

\section{Conclusions}
We have further developed the theory of Clifford codes by
giving a necessary and sufficient condition that allows to decide when
a Clifford code is equal to a stabilizer code. This condition allowed
us to compile a table of all Clifford codes that are not
stabilizer codes for error groups of order up to 255. 

One of the main open problems in the theory of Clifford codes is
whether there exists a true Clifford code which is better than any
stabilizer code. We expect the characterization shown in this paper to
be useful in the study of Clifford codes for composite systems, where
the error basis is given by the tensor product of several copies of a
fixed basis. In this case, the group is a central product of the
individual components and the challenging task is to identify suitable
normal subgroups of this central product such that the corresponding
Clifford code outperforms any stabilizer code.

\nix{
In view of fault-tolerant operations, it remains an open
problem as to whether Clifford codes will lead to schemes with a large
group of transversally implementable operations. 
}

\paragraph{Acknowledgments.} 
This research was support by NSF grant EIA 0218582 and Texas A\&M
TITF. M.~R. acknowledges support by CFI, ORDFC, and MITACS.


\begin{thebibliography}{10}

\bibitem{ashikhmin01}
A.~Ashikhmin and E.~Knill.
\newblock Nonbinary quantum stabilizer codes.
\newblock {\em IEEE Trans. Inform. Theory}, 47(7):3065--3072, 2001.

\bibitem{bierbrauer00}
J.~Bierbrauer and Y.~Edel.
\newblock Quantum twisted codes.
\newblock {\em J. Comb. Designs}, 8:174--188, 2000.

\bibitem{magma}
W.~Bosma, J.J. Cannon, and C.~Playoust.
\newblock The {M}agma algebra system~{I}: {T}he user language.
\newblock {\em J. Symb. Comp.}, 24:235--266, 1997.

\bibitem{calderbank98}
A.R. Calderbank, E.M. Rains, P.W. Shor, and N.J.A. Sloane.
\newblock Quantum error correction via codes over {GF}(4).
\newblock {\em IEEE Trans. Inform. Theory}, 44:1369--1387, 1998.

\bibitem{feng02}
K.~Feng.
\newblock Quantum error-correcting codes.
\newblock In {\em Coding Theory and Cryptology}, pages 91--142. World
  Scientific, 2002.

\bibitem{gottesman96}
D.~Gottesman.
\newblock Class of quantum error-correcting codes saturating the quantum
  {H}amming bound.
\newblock {\em Phys. Rev. A}, 54:1862--1868, 1996.

\bibitem{klappenecker034}
A.~Klappenecker and M.~R{\"otteler}.
\newblock Beyond stabilizer codes {I}: {N}ice error bases.
\newblock {\em IEEE Transaction on Information Theory}, 48(8):2392--2395, 2002.

\bibitem{klappenecker033}
A.~Klappenecker and M.~R{\"otteler}.
\newblock Beyond stabilizer codes {II}: {C}lifford codes.
\newblock {\em IEEE Transaction on Information Theory}, 48(8):2396--2399, 2002.

\bibitem{klappenecker031}
A.~Klappenecker and M.~R{\"o}tteler.
\newblock Clifford codes.
\newblock In R.~Brylinski and G.~Chen, editors, {\em Mathematics of Quantum
  Computing}, pages 253--273. Chapman \& Hall/CRC Press, 2002.

\bibitem{knill96b}
E.~Knill.
\newblock Group representations, error bases and quantum codes.
\newblock Los Alamos National Laboratory Report LAUR-96-2807, 1996.

\bibitem{matsumoto00}
R.~Matsumoto and T.~Uyematsu.
\newblock Constructing quantum error correcting codes for $p^m$-state systems
  from classical error correcting codes.
\newblock {\em IEICE Trans. Fundamentals}, E83-A(10):1878--1883, 2000.

\bibitem{gap}
The~{GAP} {T}eam.
\newblock {\sf GAP} -- {G}roups, {A}lgorithms, and {P}rogramming.
\newblock Lehrstuhl {D} f{\"u}r {M}athematik, {RWTH} {A}achen and School of
  Mathematical and Computational Sciences, Univ. St. Andrews, Scotland, 1997.

\end{thebibliography}

\end{document}